\documentclass[epj]{svjour}
\usepackage{graphics}
\usepackage{cite}
\newcommand{\gtorder}{\mathrel{\raise.3ex\hbox{$>$}\mkern-14mu
            \lower0.6ex\hbox{$\sim$}}}
\newcommand{\ltorder}{\mathrel{\raise.3ex\hbox{$<$}\mkern-14mu
            \lower0.6ex\hbox{$\sim$}}}

\begin{document}
\title{Observational Constraints on Neutron Star Masses and Radii}
\author{M. Coleman Miller\inst{1} \and Frederick K. Lamb\inst{2}
}                     
\institute{Department of Astronomy and Joint Space-Science Institute, University of Maryland, College Park, MD 20742-2421 USA; miller@astro.umd.edu \and Center for Theoretical Astrophysics and Department of Physics, University of Illinois at Urbana-Champaign, 1110 West Green Street, Urbana, IL 61801-3080, USA; and Department of Astronomy, University of Illinois at Urbana-Champaign, 1002 West Green Street, Urbana, IL 61801-3074, USA}
%
%
\abstract{
Precise and reliable measurements of the masses and radii of neutron stars with a variety of masses would provide valuable guidance for improving models of the properties of cold matter with densities above the saturation density of nuclear matter.  Several different approaches for measuring the masses and radii of neutron stars have been tried or proposed, including analyzing the X-ray fluxes and spectra of the emission from neutron stars in quiescent low-mass X-ray binary systems and thermonuclear burst sources; fitting the energy-dependent X-ray waveforms of rotation-powered millisecond pulsars, burst oscillations with millisecond periods, and accretion-powered millisecond pulsars; and modeling the gravitational radiation waveforms of coalescing double neutron star and neutron star -- black hole binary systems.  We describe the strengths and weaknesses of these approaches, most of which currently have substantial systematic errors, and discuss the prospects for decreasing the systematic errors in each method.
} 
\maketitle
\section{Introduction}

Neutron star cores have isospin asymmetries and densities of cold matter that are substantially higher than those accessible in terrestrial laboratories. Consequently, measurements of the properties of neutron stars can provide constraints on the physics of cold dense matter that complement the constraints obtained from nuclear physics experiments.  In particular, precise and reliable measurements of both the gravitational mass $M$ and the circumferential radius $R$ of several neutron stars with sufficiently different masses would provide valuable guidance for improving models of such matter.  Knowledge of stellar radii would be especially valuable, because the models of cold dense matter currently under discussion predict values for the radius of a $1.5~M_\odot$ star that range from $\approx\,$10~km (see, e.g., Figure~11 of \cite{2013ApJ...773...11H}) to $\approx\,15$~km (see, e.g., Figure~3 of \cite{2015A&A...577A..40B}), although most models predict radii between $\approx\,$11~km and $\approx\,$13~km (see, e.g., \cite{2012PhRvC..85c2801G}). 

This article reviews the current status of astrophysical measurements of $M$ and $R$, focusing primarily on developments within the last 18 months. For reviews of earlier work on determining $M$ and $R$, see, e.g., \cite{2013arXiv1312.0029M,2014EPJA...50...40L}. For an alternative view, see~\cite{2013RPPh...76a6901O}. For a recent review of the implications of $M$ and $R$ measurements for nuclear physics, see~\cite{2013ApJ...773...11H}.

We begin by discussing methods that measure only the masses of neutron stars and the results of such measurements, which for some systems are both precise enough to highly constrain the properties of neutron star matter and reliable enough to be used with high confidence. The highest precisely measured masses are $\approx 2~M_\odot$. Masses this high rule out many previously proposed soft equations of state. (Here ``soft'' means a relatively low pressure at a given density; stars constructed with such equations of state tend to have comparatively small maximum masses and a small radius for a given mass. A ``hard'' equation of state has the opposite properties.)

We then describe various methods that have been tried or proposed for measuring neutron star radii and the results that have been obtained to date. Some of these methods also provide simultaneous measurements of the mass of the star. For the reasons we discuss, most of the measurements that have so far been made using these methods are dominated by systematic errors and therefore cannot be used with confidence.  If, however, the systematic errors in these methods can be reduced, or proposed alternate methods that are less affected by systematic errors are successful, there are good prospects for obtaining reliable and precise $M$ and $R$ estimates by analyzing data collected by future X-ray satellite missions.

We conclude our review with brief discussions of some other possible methods for obtaining reliable mass and radius measurements. These include radio observations of pulsar precession in double neutron star systems, X-ray detection of very high neutron star spin rates, identification and analysis of quasi-periodic X-ray oscillations in the tails of magnetar superbursts, and combined analysis of gravitational-wave and electromagnetic observations of short gamma-ray bursts. 

\section{Mass Measurements}

The most precise and reliable measurements of neutron star masses have been made for neutron stars that are in a binary system with another star. This is because (1)~via gravity, mass has an effect at a distance; (2)~the underlying theory (Newton's laws for Keplerian motion and general relativity for post-Keplerian effects) is well-understood and tested; and (3)~many such systems are relatively clean, in the sense that there are no known complicating astrophysical effects that could potentially confuse or bias the dynamical mass measurement.  In some cases, measurement of post-Keplerian effects such as pericenter precession, the Shapiro delay, and orbital decay due to gravitational radiation can overdetermine the properties of the system, providing a test of general relativity.

If the orbital period $P_{\rm orb}$ of a binary system can be determined and the periodic changes in the line-of-sight velocity $K_1$ of one of the stars in the system can be measured, then to Keplerian order one can construct the mass function $f_1(M_1,M_2)$, which provides a lower limit on the mass $M_2$ of the {\it other} star. For a circular orbit,
\begin{equation}
M_2\geq f_1(M_1,M_2)\equiv{M_2^3\sin^3 i\over{(M_1+M_2)^2}}={K_1^3P_{\rm orb}\over{2\pi G}}\; .
\end{equation}
Here $M_1$ is the mass of the star whose velocity is measured and $i$ is the inclination of the orbit; $i=0$ means that we are viewing the orbit face-on, whereas $i=90^\circ$ means that we are viewing the orbit edge-on.  $M_1=0$ and $i=90^\circ$ would yield $M_2=f_1$; because $M_1$ is $>0$, $M_2$ must be $>f_1$, even if $i=90^\circ$.

To uniquely determine the masses of both stars in a binary system with a known mass function, at least two additional properties of the system must be measured. The ideal systems for such measurements are double neutron star binaries, because both objects in the binary are effectively point masses relative to their separation (a typical separation is $\sim 10^{11}~{\rm cm} \gg R \approx 10^{6}~{\rm cm}$).  If periodic pulses can be detected from at least one of the neutron stars, the high precision of pulsar timing can be brought to bear, and at least some of the post-Keplerian parameters mentioned above can be measured and used to remove degeneracies (see \cite{2009arXiv0907.3219F} for a lucid explanation of this approach).  

The two highest neutron star masses that have been precisely measured were determined in different ways.  The mass of the pulsar PSR~J1614$-$2230, which has a half-solar-mass white dwarf companion, was determined by measuring the mass function of the system and the Shapiro delay using radio observations of the pulsar~\cite{2010Natur.467.1081D}.  The Shapiro delay is a relativistic effect that causes the light-travel time through the gravitational well of a star to be greater than in the Newtonian limit and to vary periodically with the orbital phase of the system relative to our line of sight to the system. The magnitude of the delay depends on the mass of the pulsar's companion, while its variation depends on the inclination of the system relative to our line of sight (e.g., if the system is face-on to us, the delay has no orbital-phase dependence). Importantly, the Shapiro delay does not depend on the {\it nature} of the companion. It is therefore irrelevant whether the pulsar's companion is a neutron star, a white dwarf, or a main sequence star.  Measuring the Shapiro delay determines the two additional system parameters needed to obtain a unique solution for the masses of both stars in the system.  The estimated mass of PSR~J1614$-$2230 is $1.97\pm 0.04~M_\odot$ \cite{2010Natur.467.1081D}.

PSR~J0348$+$0432 has a white dwarf companion with observable atmospheric spectral lines~\cite{2013Sci...340..448A}.  The periodic variation of the energies of these lines yields a second mass function, while the measured gravitational redshift of the lines can be used to determine the white dwarf mass, closing the system of equations.  The estimated mass of PSR~J0348$+$0432 is $2.01\pm 0.04~M_\odot$~\cite{2013Sci...340..448A}.  

The improved timing techniques that have been developed for pulsar timing arrays are yielding more precise masses for many stars~\cite{2016MNRAS.455.1751R,2016arXiv160300545F}, so there is hope that neutron stars with even higher masses will eventually be discovered.

Non-dynamical measurements of neutron star masses are much more subject to complicating astrophysical effects that can introduce significant systematic errors.  There are, however, hints that individual ``black widow'' pulsars (so-called because the relativistic wind from the pulsar strips mass from its binary companion) may have relatively high masses.  Estimates of the mass of PSR~B1757$+$20 yield a best value of $2.40~M_\odot$ with a statistical uncertainty of $\pm 0.12~M_\odot$, but the systematic uncertainty in the system inclination allows a mass as low as $1.66~M_\odot$~\cite{2011ApJ...728...95V}.  The estimate of the mass of PSR~J1311$-$3430 obtained by fitting a basic (quiescent) light curve model to the data is $2.68\pm0.14~M_\odot$~\cite{2012ApJ...760L..36R}, but the best-fit basic model has significant, systematically varying residuals and provides a poor fit to the data, indicating the presence of unmodeled obscuration or emission~\cite{2012ApJ...760L..36R}.  The light curve and spectral data indicate the presence of emission---probably produced by heating of gas in the system by intrabinary shocks and indirect heating of the companion star---that is not included in basic light curve models. If such excess emission is present, fits of basic light curve models tend to give estimates of the pulsar's mass that are higher than the actual mass. Analyses that try to take the excess emission into account have found that the mass of PSR~J1311$-$3430 could be as low as $1.8~M_\odot$~\cite{2015ApJ...804..115R}, while the mass of PSR~J2215+5135, which was originally thought to be $>1.75~M_\odot$ at the $3\sigma$ level~\cite{2014ApJ...793...78S}, is more likely $\sim\,$1.6~$M_\odot$~\cite{2015ApJ...809L..10R}. In both cases the best-fit light curve model gives a poor fit to the data, with systematically varying residuals. Improved models and more precise observations will be required to accurately and reliably determine the masses of the neutron stars in these systems.

\section{Radius Measurements Using the Fluxes and Spectra of Thermal Emission from Non-Accreting Neutron Stars}

All current methods for determining the radii of neutron stars using their X-ray fluxes and spectra are subject to astrophysical effects that can confuse or bias the radius measurement.  Furthermore, in most cases the data are not yet precise enough to determine whether the model being used correctly describes the data.  It is therefore possible that a model may provide a statistically good fit and an apparently tight radius constraint but a value for the radius that is strongly biased relative to the true value.

Two types of cool neutron stars are of special interest for determining neutron star radii. One type is found in the quiescent low-mass X-ray binaries (qLMXBs), so named because they are thought to be binary star systems that contain a neutron star and another star that is not currently donating mass to the neutron star but has done so in the past (for reviews, see~\cite{2004NuPhS.132..427C,2004PhDT........11H}). Another type of interest is the isolated cooling neutron stars (for a review, see~\cite{2011ASSP...21..345M}).

It is only possible to reliably determine the radius of neutron stars that have atmospheric temperatures low enough that absorption processes are significant if the composition and other properties of the atmosphere are known. This is illustrated by the fact that fits of nonmagnetic helium model atmospheres to the spectra of cooling neutron stars in qLMXBs commonly yield radius estimates \mbox{$\sim\,$50\%} larger than those obtained by fitting nonmagnetic hydrogen model atmospheres to the same data (see~\cite{2012MNRAS.423.1556S,2013ApJ...764..145C,2014MNRAS.444..443H}). Similarly, hydrogen, helium, and carbon model atmospheres (whether magnetic or nonmagnetic) often give equally good fits to the spectra of isolated cooling neutron stars, but very different radii (see, e.g.,~\cite{2015A&A...573A..53K} for hydrogen and carbon fits to the spectrum of the isolated cooling neutron star in HESS~J1731$-$347). 

When model atmospheres with different compositions give different radii but equally good fits, additional information is required to determine the correct atmospheric model and hence the correct radius estimate. If the neutron star is in a binary system and has accreted matter from its companion, and if a clean measurement of the companion's spectrum---and hence the composition of its atmosphere---is available, or the binary system is ultracompact, implying that a hydrogen-rich companion could not fit within the orbit, it may be possible to determine the composition of the accreted atmosphere of the neutron star accurately enough to make a reliable estimate of the star's radius.

Fits of the spectra predicted by nonmagnetic hydrogen model atmospheres to the observed spectra of some qLMXBs have yielded $\sim\,$9--10~km radius estimates (see, e.g., \cite{2013ApJ...772....7G}). However, several authors have cautioned that these estimates may be affected by systematic errors. In addition to any uncertainty in the composition of the neutron star's atmospheres, any errors in the assumed interstellar absorption along the line of sight will affect the radius estimate~\cite{2014MNRAS.444..443H}. Because qLMXBs have comparatively low temperatures, they emit mostly soft X-rays; consequently, their observed spectra are strongly affected by passage through the interstellar medium. Different models of the interstellar absorption can therefore produce significantly different radius estimates and uncertainties (see, for example, Figure~6 of \cite{2014MNRAS.444..443H}). For this reason, qLMXBs with low absorbing columns (e.g., the qLMXBs in 47~Tuc \cite{2003ApJ...588..452H,2006ApJ...644.1090H}) are preferred for neutron star radius estimates.

In principle, qLMXBs in globular clusters are the best ones to use for radius measurements, because their distances can be determined independently of their measured fluxes, but the compositions of the atmospheres of these qLMXBs are typically poorly known. 

Although the spectra of most qLMXBs do not have a time-varying power-law component~\cite{2015MNRAS.452.3475B}, some do (see, e.g., \cite{2013ApJ...772....7G,2014MNRAS.444..443H}). There is some possibility that the varying power-law spectral component seen in the latter systems is produced by {\it nonthermal} emission, raising the question whether some or even most of the radiation from these stars is produced by a process other than thermal emission from the stellar surface.  If, for example, some of the observed emission is produced by residual accretion, it could well be that it comes from only a fraction of the full stellar surface. If in addition the center of the emitting region is moderately close to the rotational axis (see, e.g., \cite{2009ApJ...706..417L,2009ApJ...705L..36L} for arguments that the observed properties of many actively accreting LMXBs point to this accretion geometry), there would be no telltale pulsations. Even a very weak stellar magnetic field could channel matter accreting at low rates onto an area that covers less than the full stellar surface. If this is the case, the area inferred by measuring the flux and spectrum would be smaller than the full surface area of the star. 

Similar concerns apply if a significant fraction of the radiated energy is being supplied by relativistic magnetospheric return currents entering the stellar surface. Unlike accretion, these are expected to produce emission that is almost entirely thermal, because the column depth at which the relativistic electrons deposit their energy is expected to be much larger than the column depth at which the optical depth is unity (see, e.g., \cite{1974RvMP...46..815T}).

The bounds on the systematic uncertainties in radius estimates that are implied by the absence of detected pulsations are currently being investigated by K.~Elshamouty and colleagues (in preparation). As usual, better observations (e.g., from the approved mission {\it Athena}; see \cite{2013arXiv1306.2334M}) and better modeling are likely to be the keys to progress using this radius estimation method. However, the low count rates produced by non-accreting neutron stars and the lack of features in their spectra pose substantial challenges.

\section{Mass and Radius Measurements Using the Fluxes and Spectra of Thermonuclear X-ray Bursts}

The approach to estimating neutron star radii that has recently received the most attention uses the measured fluxes and spectra of thermonuclear X-ray bursts. This method can also provide a simultaneous estimate of the star's mass. The mass and radius estimates derived using this approach have been the subject of significant controversy. We therefore describe the different variants of this approach and their relative strengths and weaknesses in some depth. We begin by providing background information about X-ray bursts and their spectra and then discuss the methods that have been used. We describe the surprisingly large differences between the results obtained by analyzing the same data using frequentist and Bayesian methods and the constraints on $M$ and $R$ that have been derived using measurements of burst fluxes and spectra. Finally, we discuss the inconsistencies between the simplified model currently being used and most of the burst data that have been analyzed to date.

\subsection{Background}

Thermonuclear bursts are produced by runaway nuclear burning of matter that has accreted onto the surface of a neutron star in a close binary system;. (See \cite{1976ApJ...205L.127G,1976ApJ...206L.135B,1976Natur.263..101W,1977Natur.270..310J,1978ApJ...220..291L} for selected early observational and theoretical papers about these bursts and \cite{2006csxs.book..113S} for a relatively recent review about bursts that incorporates what we have learned using data from the {\it Rossi} X-ray Timing Explorer ({\it RXTE}).)  Typical bursts last for seconds and are thought to be powered by helium fusion, but some last for tens of seconds to minutes; the latter are thought to be powered by fusion of both hydrogen and helium. Some much rarer bursts have approximately the same peak luminosity as typical bursts, but can last for hours. These ``superbursts'' are likely powered by thermonuclear fusion of carbon in much deeper layers of the star (see, e.g., \cite{2001ApJ...559L.127C}).

If the atmosphere of a neutron star undergoing a thermonuclear X-ray burst were a blackbody, its radius could be determined using its spectrum and distance. Indeed, this method can be used to successfully measure the radii of ordinary main sequence stars like the Sun that have highly absorbing atmospheres with opacities that change only gradually with photon energy. The bolometric luminosity $L$ of the star can be computed from its distance and its flux observed at Earth, and its surface temperature $T$ can be determined by fitting the Planck function to its observed spectrum. The star's radius $R$ can then be estimated using the relation $L=4\pi R^2\sigma_{\rm SB} T^4$, where $\sigma_{\rm SB}$ is the Stefan-Boltzmann constant. Radius estimates made in this way can be checked using the ten or so stars, including the Sun, whose angular radii and distances are known, or whose radii have been determined by asteroseismology. It turns out that the radius estimate made by fitting the blackbody function to the observed spectrum of these stars is fairly accurate. Besides validating the use of this method to infer the radius of such stars, this exercise shows that the atmospheres of these main sequence stars radiate approximately like blackbodies.

The shapes of the X-ray spectra observed during thermonuclear X-ray bursts are also often adequately described (i.e., within the statistical uncertainties) by the Planck function (see, e.g.,~\cite{1977ApJ...212L..73S,1993SSRv...62..223L}). However, applying the method just discussed to the fluxes and spectra of such bursts yields neutron star radii that are are unreasonably small, often \mbox{$\sim\,$5~km} or even less (see, e.g.~\cite{1979A&A....78L..15G,1993SSRv...62..223L}).  It also yields fluxes much greater than the Eddington critical flux $F_E$ at which the outward force per unit mass caused by interaction of the radiation with the gas in the neutron star atmosphere balances the inward gravitational acceleration, which is inconsistent with models of these atmospheres (see, e.g.~\cite{1982ApJ...260..815M,1993SSRv...62..223L}). The problem is that although the {\it shape} of the observed spectrum is fairly well described by the shape of the Planck function, the {\it normalization} is not the Planck normalization, because the atmosphere of a neutron star is far from being a blackbody (see Section~5 of~\cite{1993SSRv...62..223L}).

At the temperatures $\sim\,$10$^7$~K that are characteristic of the surfaces of neutron stars during thermonuclear X-ray bursts, the opacity in the atmosphere is dominated by electron scattering rather than absorption; this is especially true if the atmosphere is mostly hydrogen or helium, because these elements will be fully ionized. The backscattering caused by electron scattering reduces the radiative efficiency of the atmosphere. The spectrum of the radiation from such an atmosphere has approximately the same shape as a blackbody spectrum, but the flux it radiates at a given temperature is substantially less than the flux radiated by a blackbody with the same temperature. For fluxes greater than $\sim\,$20\% of the Eddington flux, the shapes of burst spectra are more accurately described by Bose-Einstein spectra with a chemical potential $\mu \approx -kT$~\cite{2010ApJ...720L..15B,2011fxts.confE..24M,2013IAUS..290..101M}; these are the thermodynamic equilibrium spectra produced by scattering when the number of photons is conserved~\cite{1975SovAstron.18..413I}. For early detailed model atmosphere computations of neutron star spectra, see~\cite{1982A&A...107...51V,1984ApJ...287L..27L,1986ApJ...306..170L,1986PAZh...12..918L,1986ApJ...305L..67E,1987PASJ...39..287E,1988AdSpR...8..477K,1989ApJ...339..386M,1991MNRAS.253..193P,1991AcA....41...73M,1991ApJ...376..161M,1991ApJ...377L..93L,1997A&A...320..177M,2004ApJ...602..904M,2005A&A...430..643M} and Section~5 of~\cite{1993SSRv...62..223L}.  More recent model atmosphere computations~\cite{2008A&A...490.1127R,2011A&A...527A.139S,2012A&A...545A.120S,2015arXiv151006962S} are more accurate and have been validated to very high accuracy and precision by comparing them with spectral data from a superburst from 4U~1820$-$30~\cite{2011fxts.confE..24M,2013IAUS..290..101M}.

The local bolometric radiative energy flux from a stellar surface can be described by the effective temperature $T_{\rm eff}$, defined implicitly by the relation $F = \sigma_{\rm SB} T_{\rm eff}^4$. If the surface were a blackbody, $T_{\rm eff}$ would be equal to the kinetic temperature of the surface. A useful way of characterizing burst spectra is by the temperature $T_c$ obtained by fitting a Planck function to the spectrum observed over a wide energy band. Following~\cite{1979rpa..book.....R}, we refer to $T_c$ as the color temperature. In the X-ray burst literature, $T_c$ is sometimes called the blackbody temperature, but this could be misleading because the neutron star surface is not a blackbody and does not emit a blackbody spectrum. The ratio $f_c \equiv T_c/T_{\rm eff}$ is called the color factor. A useful approximate  expression for the specific intensity of the radiation from a burst atmosphere is $I\approx f_c^{-4}I_{\rm BB}(f_cT_{\rm eff})$, where $I_{\rm BB}(T)$ is the blackbody specific intensity at temperature $T$.  

As noted above, a scattering-dominated atmosphere is an inefficient radiator, so its temperature must be higher than the temperature of a blackbody that radiates the same flux. Consequently, $T_c$ will be higher than $T_{\rm eff}$. Theoretical computations of X-ray burst spectra that have been validated by comparing them with highly accurate measurements of burst spectra show that $T_c$ is typically $\sim\,$40--90\% higher than $T_{\rm eff}$ (i.e., $f_c$ is 1.4--1.9) for the X-ray fluxes of interest \cite{2012A&A...545A.120S}.

For examples of fits of blackbody and model atmosphere spectra to the observed X-ray spectra of neutron stars that are not accreting significantly at the time of observation, see \cite{1999ApJ...514..945R,2000A&A...358..583C,2001ApJ...551..921R,2001ApJ...559.1054R}.

Some of the methods we discuss for deriving neutron star masses and radii from measurements of burst fluxes and spectra use data from so-called photospheric radius expansion (PRE) bursts. The nature of these bursts and the definitions of their phases will be important for the subsequent discussion, so we describe them here.

The observational signature of a PRE burst is the occurrence of two maxima in the temperature of the best-fit Planck spectral shape as the burst evolves. In a typical PRE burst, the temperature first increases (the burst ``rise''), then dips, increases again, and finally decreases more slowly. The initial temperature rise occurs when energy from the layer heated by thermonuclear burning first reaches the stellar surface. The temperature subsequently dips when the radiative energy flux through the atmosphere at the stellar surface temporarily exceeds the Eddington flux. When this happens, the ionized gas near the stellar surface is driven outward, creating a quasi-static low-density expanded atmosphere if the radiative flux is low enough that the atmosphere can achieve equilibrium by expanding (the Eddington flux increases with increasing radius). If instead the flux through the atmosphere is too high for the atmosphere to achieve equilibrium by expanding, the atmosphere will continue to expand, creating a wind.

Two surfaces in the burst atmosphere or wind are particularly relevant to their dynamics and the radiation flux and spectrum they produce. Assume for simplicity that the atmosphere is roughly spherically symmetric (if it is not, similar physics applies, but the situation is more complicated). One important surface is the surface of last scattering, i.e., the surface from which photons propagate unimpeded into the surrounding space. The radius of this surface is the largest radius at which the momentum of the photons couples to the gas. Because the Eddington flux increases with increasing radius, it is the Eddington flux at the surface of last scattering that determines the momentum (and energy) flux of the radiation leaving the atmosphere or wind (dynamical calculations~\cite{1982ApJ...258..696W,1983ApJ...267..315P,1983PASJ...35...17E,1983PASJ...35...33K,1985ApJ...289..634Q,1986ApJ...302..519P,1987ApJ...312..700J,1994ApJ...433..276N} show that the radiative flux from winds is only slightly greater than the Eddington flux computed at large radii).

A second important surface is the surface where photons are last thermalized by the hot gas in the atmosphere or wind. The kinetic temperature at this surface determines the spectrum of the radiation that leaves the atmosphere or wind. In a scattering-dominated atmosphere or wind, this surface lies inside the surface of last scattering. If the density scale height in the outer atmosphere or wind is comparable to the radius, the radius $R_{\rm th}$ of the thermalization surface may be substantially smaller that the radius $R_{\rm sc}$ of the surface of last scattering. If instead the density scale height is much smaller than the radius, $R_{\rm th}$ will be almost the same as $R_{\rm sc}$.

Most of the X-ray burst literature does not distinguish the thermalization surface from the surface of last scattering, implicitly assuming they are one and the same. This ``combined'' surface is called the photosphere. Not distinguishing these two surfaces usually does not introduce significant errors into the analysis when the atmosphere is quiescent, because the density scale height of a quiescent neutron star atmosphere is very small compared to the star's radius, but it can introduce substantial errors if the atmosphere is expanded, as it is during portions of PRE bursts. In order to connect with the existing literature, we will refer to the combined thermalization and last scattering surfaces as the photosphere, but will distinguish them when their difference is likely to be important.

As the thermalization radius increases during a PRE burst, the temperature of the radiation escaping from the atmosphere drops. If it falls far enough, the energy flux in the detector bandpass will decrease. As the star continues to radiate, the layer that was heated by thermonuclear burning cools, causing the outward energy flux from the heated layer to diminish. When the outward energy flux becomes small enough that the outward radiative acceleration in the outer atmosphere is less than the inward gravitational acceleration there, the gas there begins to settle inward. 

As the atmosphere settles, the last scattering and thermalization radii decrease until they are again very close to the radius given by stellar structure calculations. As the thermalization radius decreases, the temperature of the radiation escaping from the atmosphere increases until the thermalization radius reaches the quiescent stellar surface (this event is called ``touchdown''~\cite{1990A&A...237..103D}). At this moment, the outward radiative flux at the stellar surface is exactly equal to the local Eddington flux, and the color temperature of the radiation escaping from the atmosphere is expected to have a local maximum. Following touchdown, the temperature of the radiation slowly decreases as the outer layers of the star cool. This last phase lasts seconds to tens of seconds and is referred to as the ``cooling phase'' or ``cooling tail'' of the burst.

In the following sections we refer to the bolometric flux at various times during X-ray bursts. Because X-ray satellites measure fluxes only in a finite energy band, bolometric fluxes must be determined by extrapolating the measured flux to all energies, using a model of the observed spectrum. At and after touchdown, burst spectra peak at a few keV, well within the bandpasses of the detectors on X-ray timing satellites such as {\it EXOSAT} and {\it RXTE}, and are well described by a Planck function. Consequently, the bolometric flux during these phases has been determined accurately enough that the effects of any resulting systematic errors are small compared to other potential systematic errors in analyzing bursts. It is much more difficult to determine the bolometric flux when the burst atmosphere is expanded or there is a wind, because the resulting spectra can peak at much lower energies, where interstellar absorption is important and X-ray satellites such as {\it EXOSAT} and {\it RXTE} had little or no sensitivity.

\subsection{Methods}

The observed properties of thermonuclear X-ray bursts have been used by many authors to estimate the masses and/or radii of neutron stars (see~\cite{1979ApJ...234..609V,1981ApJ...247..628H,1982A&A...107...51V,1986ApJ...305..246F,1987MNRAS.226...39S,1987A&A...172L..20V,1988AdSpR...8..477K,1990A&A...237..103D,1990PASJ...42..633V,1990Ap&SS.173..171K,2006Natur.441.1115O,2009ApJ...693.1775O,2010ApJ...712..964G,2010ApJ...719.1807G,2010ApJ...722...33S,2011A&A...527A.139S,2011ApJ...742..122S,2012A&A...545A.120S,2012ApJ...747...77G,2012ApJ...747...76G,2012ApJ...748....5O,2013IAUS..291..145P,2013ApJ...765L...1G,2013ApJ...765L...5S,2014MNRAS.442.3777P,2015arXiv150505155O,2015arXiv150902924O,2015arXiv150906561N,2015arXiv151006962S}).

Of special note are the papers~\cite{1986ApJ...305..246F,1987MNRAS.226...39S,1987A&A...172L..20V,1990A&A...237..103D,1990PASJ...42..633V} that developed and demonstrated the X-ray flux--color temperature method that is now used with some variations by most researchers. These papers derived and applied the two relations between $M$ and $R$ specified, respectively, by (1)~the X-ray luminosity during the contraction of the photosphere of a PRE burst, assuming the luminosity during this phase of the burst is the Eddington luminosity at the photosphere, and (2)~the luminosity and color temperature during the cooling phase of the burst, assuming the emission during this phase comes from the quiescent stellar surface. This system of equations was closed to obtain $M$ and $R$ estimates by using either (a)~the distance to the star to convert the luminosities in these relations to observed bolometric fluxes or (b)~the gravitational redshift from the quiescent stellar surface to the observer, to determine $M/R$ independently of relations~(1) and~(2). The gravitational redshift from the surface was estimated either (i)~by measuring the energy of an identified atomic spectral line originating at the stellar surface, if one was observed, or (ii)~by measuring the variation of the Eddington luminosity, as measured by a distant observer, at different values of the photospheric radius during the contraction phase of one or more PRE bursts from the star. At present there are no confirmed detections of atomic spectral lines originating at the surfaces of neutron stars, and hence no secure gravitational redshifts have been derived from such lines.

In~\cite{1986ApJ...305..246F} (see Figure 4.6 of~\cite{1993SSRv...62..223L} for a corrected version of Figure~2) the mass and radius of 4U~1636$-$536 were estimated using variants of relations~(1) and~(2) and closing the system of equations using the surface redshift given by the energy of an observed spectral line (it is now generally agreed that this line was incorrectly identified). The approach developed in~\cite{1987MNRAS.226...39S} to estimate the mass and radius of 4U~1746$-$37 is very similar to the approach currently being used by most researchers. This analysis used relations~(1) and~(2) described above, determining the Eddington luminosity of the star by assuming it is equal to the luminosity observed during the phase of the burst when the observed bolometric flux is at its maximum and constant. This approach implicitly assumes that the thermalization radius is the same as the radius of last scattering. The system of equations was closed using an estimate of the distance to the star. The analysis explicitly included the possibility that burst radiation may come from a fraction of the full stellar surface and could be beamed or obscured. Reference \cite{1987MNRAS.226...39S} cites theoretical analyses which find that the color factor $f_c$ should vary during the cooling phase of bursts and takes this variation into account. It also considers the possibility that the emission observed during the burst could be contaminated by persistent emission. 

In~\cite{1987A&A...172L..20V}, this same method was used to estimate the mass and radius of 4U~1820$-$30, which is in the globular cluster NGC~6624. In~\cite{1990A&A...237..103D}, PRE bursts from four neutron stars were analyzed to determine the surface redshifts of these stars. This was done using relation~(1) evaluated when the atmosphere is expanded and at the moment of touchdown. The analysis of the expanded atmosphere implicitly assumes that the thermalization radius is the same as the radius of last scattering. An important point made in this paper and subsequently incorporated into the standard analysis method is that the luminosity observed at touchdown is exactly the Eddington luminosity at the stellar surface, appropriately redshifted. In~\cite{1990PASJ...42..633V}, the method introduced in~\cite{1987MNRAS.226...39S} was applied to 4U~2129$+$11, which is in the globular cluster M15. This paper is noteworthy because it carefully explores the systematic errors that can be caused by unmodeled variations in the color temperature, deviations of the observed spectral shape from the shape of the Planck function, and possible beaming and obscuration of the emission from the star, and explicitly includes the Klein-Nishina reduction in the scattering opacity at large photon energies.

Current efforts to determine $M$ and $R$ using the observed properties of thermonuclear X-ray bursts benefit enormously from the superb quality of the burst data that have been collected using {\it RXTE}. The general approach currently being used by most researchers uses two relationships between $M$ and $R$ that are closely related to relations~(1) and~(2) discussed above. One equation relates $M$ and $R$ to the bolometric flux observed at the moment of touchdown. The assumption made in deriving this relation is that the luminosity at the stellar surface at touchdown is the local Eddington luminosity there. This local luminosity is converted to the observed bolometric flux using the distance and the redshift of the stellar surface. The moment of touchdown is identified as the moment when the color temperature $T_c$ reaches its maximum. By using the bolometric flux and temperature at touchdown, this approach avoids the systematic error that could arise when the thermalization radius and the radius of last scattering are significantly different. A second equation relates $M$, $R$, and the color factor $f_c$ to the bolometric flux and color temperature observed during the cooling tail. This equation is evaluated using a model for how $f_c$ depends on $T_{\rm eff}$, the chemical composition of the atmosphere, the surface gravity, and other parameters. The system of two equations is closed using the distance to the star. 

Two versions of this general approach are currently in use. In one version~\cite{2006Natur.441.1115O,2009ApJ...693.1775O,2010ApJ...712..964G,2010ApJ...719.1807G,2010ApJ...722...33S,2012ApJ...747...77G,2012ApJ...747...76G,2012ApJ...748....5O,2013ApJ...765L...1G,2013ApJ...765L...5S,2015arXiv150505155O}, the bolometric flux at touchdown is measured directly. Doing this requires observing at least one PRE burst, determining the moment of touchdown during the burst, and measuring the bolometric flux at that moment.

In another version of this approach~\cite{2011A&A...527A.139S,2011ApJ...742..122S,2012A&A...545A.120S,2013IAUS..291..145P,2014MNRAS.442.3777P,2015arXiv150906561N,2015arXiv151006962S}, which the practitioners call the cooling tail method, the bolometric flux at touchdown is determined indirectly by tracking the evolution of the observed bolometric flux and color temperature and using theoretical computations of the evolution of the color factor to estimate the bolometric flux at touchdown. Strictly speaking, this method does not require observations of any PRE bursts from the star in question. What is required are observations of bursts in which the X-ray flux from the stellar surface changes enough that the color factor changes. This method can also be used to analyze PRE bursts.

Both these approaches share many common assumptions. One is that the spectral model being used to analyze the data from the cooling tail accurately describes the bursts being analyzed; if it does not, any inferences---at least using the discrepant bursts---are suspect. The X-ray spectra predicted by the most recent detailed stellar atmosphere models~\cite{2012A&A...545A.120S}, when fit to the spectral evolution of the only burst for which there is enough data to distinguish between spectral models (a superburst from 4U~1820-30), give fits near the maximum of the burst that are excellent, far superior to those obtained by fitting Planck spectra, and much superior to those obtained by fitting  Bose-Einstein spectra~\cite{2013IAUS..290..101M}. This increases confidence in the accuracy of these models and inferences made using them.

It is disquieting, however, that most bursts do not follow the evolution predicted by these spectral models~\cite{2010ApJ...722...33S,2011ApJ...742..122S,2014MNRAS.445.4218K}. This likely indicates that during most bursts, emission from sources other than the stellar surface contributes to the bolometric flux and/or affects the observed spectrum. Possible sources could include a boundary layer between the disk and stellar surface, or radiation reflected from the accretion disk.

The approaches currently being used also assume that the entire stellar surface emits uniformly, both at touchdown and during the cooling tail, at least until persistent emission is again an important contributor to the observed flux and spectrum. If this assumption is not valid, making it introduces a systematic error in the measurements of $M$ and $R$. The presence of brightness oscillations at the stellar rotational frequency during many bursts~\cite{2001ApJ...553L.157M,2004ApJ...608..930M,2008ApJS..179..360G,2012ARA&A..50..609W} demonstrates that, at least during these bursts, the stellar surface is not emitting uniformly.

There is also evidence for nonuniform emission from the stellar surface even in cases where no oscillations are detected. A detailed analysis of the 4U~1820-30 superburst has demonstrated that the emitting area decreases by \mbox{$\sim20$\%} over a half-hour interval following the peak of the burst~\cite{2013IAUS..290..101M}. This evidence that the emitting area becomes less than $\sim\,$80\% of the stellar surface, combined with the absence of any evidence of oscillations despite the very large number of counts that were collected, implies that the emitting region is nearly symmetric around the star's spin axis. A reasonable interpretation is that the emitting region includes the spin pole. 

This interpretation is consistent with observational evidence and theoretical arguments that the persistent emission of accretion-powered millisecond X-ray pulsars is concentrated near their spin poles~\cite{2009ApJ...705L..36L,2009ApJ...706..417L}. Some of the X-ray bursts from these pulsars have detectable oscillations~\cite{2001ApJ...553L.157M,2004ApJ...608..930M,2008ApJS..179..360G,2012ARA&A..50..609W,2014ApJ...792....4C}, implying that the emission during these bursts is not symmetric around the spin axis, but others show no detectable oscillations~\cite{2001ApJ...553L.157M,2004ApJ...608..930M,2008ApJS..179..360G,2014ApJ...792....4C}, implying that the emission during these bursts {\it is} fairly symmetric around the spin axis. 

The intermittent accretion-powered millisecond X-ray pulsar HETE~J1900.1$-$2455 produces bursts with oscillations and bursts without  detectable oscillations~\cite{2009ApJ...698L.174W}. The persistent accretion-powered millisecond X-ray pulsar IGR~J17498$-$2921 also produces bursts with and without detectable oscillations~\cite{2012MNRAS.422.2351C}. The IGR~J17498$-$2921 bursts show flux differences that are due primarily to differences in the emitting area rather than the temperature, implying that some of these bursts have emitting areas that are only a fraction of the entire stellar surface~\cite{2012MNRAS.422.2351C}. There are low upper limits on the amplitudes of any oscillations during some of these bursts, indicating that the emitting area is well-centered on the rotational axis~\cite{2012MNRAS.422.2351C}.

Given these results, the absence of detectable oscillations during bursts from neutron stars is not compelling evidence that the emission during these bursts is uniform over the entire stellar surface.
If burst emission is interpreted as coming uniformly from the entire stellar surface but is actually nonuniform, this will introduce a systematic error in estimates of $R$, biasing them in the direction of smaller radii (see, e.g., \cite{1979ApJ...234..609V}).

Other assumptions currently made that may not be valid in all cases are that the burst emission is not affected by beaming (e.g., by blocking by the accretion disk), that the emission is not significantly contaminated by emission produced by continuing accretion, and that the burst spectrum is not affected by absorption, obscuration, or reflection of radiation by gas in the system, including by the disk (see, e.g.~\cite{1986A&A...157L..10V,1987MNRAS.226...39S,1987A&A...172L..20V,1990PASJ...42..633V,1993SSRv...62..223L}).

Future, more sensitive X-ray missions may be able to use other methods to constrain or determine $M$ and $R$ using observations of X-ray bursts. As one example, the {\it NICER} mission will have a detector that has high time resolution over an energy band that extends down to 0.1~keV~\cite{2012SPIE.8443E..13G}. Observations of PRE bursts using this instrument might be able to measure accurately the flux from the photosphere even when its radius is very large, allowing an accurate application of the method proposed in~\cite{1987MNRAS.226...39S}. Computations of the X-ray flux and spectrum from expanded atmospheres that take into account the difference between the thermalization radius and the radius of last scattering would be needed to determine $M/R$ with high accuracy using this method. As another example, $M$ and $R$ could be determined directly by fitting accurate model spectra to sufficiently high-resolution spectral data~\cite{Majczyna:2005tx,2013IAUS..290..101M}. Only the shape of the spectrum matters in this method, not the absolute flux. Consequently, this method does not require knowledge of the distance to the star or the fraction of the stellar surface that is emitting, so long as the emitting portion has a uniform spectrum. It is even immune to photon energy-independent shifts in the calibration of the instrument (although energy-dependent shifts would affect this method).

\subsection{X-ray flux-color temperature method}

One quantity required in the X-ray flux--color temperature method currently being used is the bolometric flux observed at touchdown. Assuming that the luminosity of the stellar surface at the moment of touchdown is the local Eddington luminosity $L_{E}(R)$, this flux is~\cite{2009ApJ...693.1775O,1987MNRAS.226...39S} 
\begin{equation}
F_{TD,{\rm obs}}=\xi_F {L_{E}(R)\over{D^2}}\left(1-2\beta\right)^{1/2}
=\xi_F{GMc\over{\kappa D^2}}\left(1-2\beta\right)^{1/2}\;.
\label{eq:FTD}
\end{equation}
Here ``TD'' denotes the moment of touchdown, the subscript ``obs'' indicates that this is the flux measured by an observer far from the star and $\xi_F$ is the factor needed to correct for the effects of nonuniform emission from the stellar surface, beaming, absorption, obscuration, and reflection of the burst radiation, and contamination of the flux by other emission from the system. Current practice assumes all these effects are negligible, i.e., that $\xi_F=1$. We adopt this simplification here, for the sake of illustration. The quantity $\kappa$ is the photon-energy averaged radiative opacity, $D$ is the distance to the star, the exterior spacetime is assumed to be the Schwarzschild spacetime, and $\beta \equiv GM/Rc^2$ is the stellar compactness parameter. It is often assumed that $\kappa \approx 0.2~{\rm g~cm}^{-2}(1+X)$, which is the Thomson scattering opacity of matter with a hydrogen mass fraction $X$, but at the $\sim\,$3~keV temperatures of X-ray burst radiation, Klein-Nishina corrections reduce the scattering cross section by as much as several percent~\cite{1993SSRv...62..223L}.

The second quantity that is needed in the current X-ray flux--color temperature method is the angular area~\cite{2009ApJ...693.1775O,1987MNRAS.226...39S}\begin{equation}
A_i\equiv \xi_A {F_{{\rm obs},i}\over{\sigma_{\rm SB}T_{c,{\rm obs},i}^4}} = f_{c,i}^{-4}\left(R\over D\right)^2\left(1-2\beta\right)^{-1}  \;,
\label{eq:A}
\end{equation}
measured at appropriate times $t_i$ in the cooling tail of the burst. Here $\xi_A$ is the factor needed to correct for the effects of nonuniform emission from the stellar surface, beaming, absorption, obscuration, and reflection of the burst radiation, and contamination of the flux by other emission from the system. Current practice assumes all these effects are negligible, i.e., that $\xi_A=1$. We adopt this simplification here, for the sake of illustration. The quantities $F_{{\rm obs},i}$ and $T_{c,{\rm obs},i}$ are the bolometric flux and the color temperature observed at time $t_i$. The quantity $f_{c,i}$ is the color factor computed using a model atmosphere with the properties of the burst atmosphere at time $t_i$.

Both $F_{{\rm obs},i}$ and $T_{c,{\rm obs},i}$ decrease as the stellar atmosphere cools, but $A$ is often nearly constant once the flux has dropped below $\sim\,$50--70\% of its peak value~\cite{1977ApJ...212L..73S}. If (as is normally assumed) the atmosphere of the neutron star is composed of hydrogen or helium or a mixture of these elements, $f_c$ decreases from $\sim\,1.9$ when the surface flux is close to the Eddington flux to $\sim\,$1.4 when the flux is half the Eddington flux. If instead the atmosphere has significant metal content, $f_c$ could be as low as 1.1--1.2 at half the Eddington flux~\cite{2015A&A...581A..83N}.  

We refer to the model defined by Equations~(\ref{eq:FTD}) and~(\ref{eq:A}) with $\xi_F=1$ and $\xi_A=1$ as the simple burst cooling model.

In one version of the X-ray flux--color temperature method, $F_{TD,{\rm obs}}$ is estimated by fitting a Planck function to the X-ray spectrum in the observed energy band and then integrating the Planck function to obtain the bolometric flux. The accuracy of this method for determining $F_{TD,{\rm obs}}$ depends only on the accuracy with which the shape of the Planck function matches the shape of the observed spectrum. In practice, the Planck function matches the observed spectrum within the detector bandpass to high accuracy, and the 2-60~keV bandpass of the {\it RXTE} PCA captures more than 90\% of the flux if $T_c \ge 3$~keV.  In the tail, $f_c$ is assumed to remain constant once the flux has dropped below $\sim\,$50--70\% of its peak value, and Equation~(3) is solved using this value of $f_c$ and the (approximately constant) value of $A$ at this time.

When estimating $M$ and $R$, it is convenient to transform from $A$ and $F_{TD}$ to the dimensionless quantities~\cite{2010ApJ...722...33S}
\begin{equation}
\alpha\equiv {F_{TD,{\rm obs}}\over{A^{1/2}}}{\kappa D\over{c^3 f_c^2}}
\label{eq:def-alpha}
\end{equation}
and
\begin{equation}
\gamma\equiv {Ac^3f_c^4\over{F_{TD,{\rm obs}}\kappa}}\; .
\label{eq:def-gamma}
\end{equation}
Note that the definitions of $\alpha$ and $\gamma$ are nonlocal in time, in the sense that each involves quantities that are measured at touchdown and late in the cooling tail. These combinations of observable quantities are related to the compactness and radius of the star by
\begin{equation}
\alpha=\beta(1-2\beta)
\label{eq:alpha}
\end{equation}
and
\begin{equation}
\gamma={R\over{\beta(1-2\beta)^{3/2}}}\; .
\label{eq:gamma}
\end{equation}
Solving Equations~(\ref{eq:alpha}) and~(\ref{eq:gamma}) for $\beta$, $R$, and $M$ yields
\begin{equation}
\beta={1\over 4}\pm{1\over 4}\sqrt{1-8\alpha}\; ,
\label{eq:beta}
\end{equation}
\begin{equation}
R=\alpha\gamma\sqrt{1-2\beta}\; ,
\label{eq:R}
\end{equation}
and
\begin{equation}
M={\beta Rc^2\over G}\; .
\label{eq:M}
\end{equation}
In this simple model, $R$ and $M$ can be estimated by inserting measured values of $\alpha$ and $\gamma$ into Equations~(\ref{eq:beta})--(\ref{eq:M}).

In the cooling tail version of the X-ray flux--color temperature method, which can be used to analyze PRE bursts but does not require them, the approach is slightly different (see~\cite{2015arXiv150906561N} for a recent application of this method to X-ray burst data). In this approach, the information from the whole cooling track after the peak of the burst is used and the observed cooling is compared to the theoretical evolution predicted by models of passively cooling neutron star atmospheres. Specifically, $F_{\rm obs}$ and $T_{c,{\rm obs}}$ are measured during the tail evolution of the burst.  For that portion of the burst when the area of the photosphere is assumed to be the area of the quiescent stellar surface (i.e., after touchdown for a PRE burst, or after the flux maximum for a non-PRE burst), $A$ should depend only on $f_c$. Because $f_c$ depends on the ratio of the local surface flux to the local Eddington flux as well as the composition and surface gravity (see, e.g., Figure~10 of \cite{2015arXiv151006962S}), model atmosphere spectra can be used to infer $F_{TD,{\rm obs}}$ and solve for $M$ and $R$ even if the burst is not a PRE burst or if it is difficult to measure $F_{TD,{\rm obs}}$ at touchdown. Figure~16(b) of \cite{2015arXiv151006962S} shows an example of this kind of fit, to data from 4U~1608$-$52. 

If the star rotates at several hundred Hertz, $A$ is modified by the rotational flattening of the star. V.~Suleimanov (private communication) has noted that rapid rotation also affects the touchdown flux because rotation changes the local net surface gravity. More generally, V.~Suleimanov (private communication) finds that the {\it sign} of the correction depends on the assumed properties of the local surface emission. If the locally measured temperature is assumed to be uniform over the surface, rotation increases the estimated radius, compared to its value for a nonrotating star~\cite{1987PASJ...39..475A,2015ApJ...799...22B,2015arXiv150505155O}; if instead the locally measured ratio of the flux to the Eddington flux is assumed to be uniform over the surface, rotation can reduce the estimated radius.

\subsection{Peculiarities of the simple burst cooling model}

The simplified model of the evolution of the flux and color temperature in the cooling tails of X-ray bursts that was described in the previous section has peculiarities that could potentially produce misleading results. Comparing it with data also provides a rare example of a stark contrast between the results given by frequentist and Bayesian analyses of a measurement problem.

If the observed flux and color temperature give $\alpha>1/8$, the model yields {\it no} solution (see~\cite{2010ApJ...722...33S}). If, on the other hand, the observations give $\alpha<1/8$, the model yields {\it two} solutions: in addition to the true $M$--$R$ pair, there is a ``ghost'' $M$--$R$ pair that gives exactly the same values of the observables $A$ and $F_{TD,{\rm obs}}$ as does the correct pair. Equation~(\ref{eq:beta}) shows that the true and ghost solutions are reflections of each other across the $\beta=1/4$ line in the $M$--$R$ plane: one is more compact and the other less compact than $\beta=1/4$, by exactly the same amount.  

The effect of the ghost solution on estimates of $M$ and $R$ depends on whether the data are analyzed using a frequentist approach or a Bayesian approach.  The key difference for our purposes here is that in the frequentist analysis the data are treated as random variables, whereas in a Bayesian analysis they are not.  In the typical frequentist approach to the present problem, quantities such as $F_{TD,{\rm obs}}$, $A$, and $D$, which are inferred from observations, and quantities such as $f_c$ and $\kappa$, which are largely derived from theoretical considerations, are treated as having error bars and thus in some sense they can be considered to define a probability space.  The portions of this probability space that yield $\alpha>1/8$ (as defined in Equation (4)) would imply complex values of $M$ and $R$ and are therefore excluded from the analysis (see \cite{2010ApJ...722...33S} for a discussion of how this can lead to misleadingly small estimates of the uncertainties in $M$ and $R$).  

In contrast, in a Bayesian approach (which was first described in this context in \cite{2015ApJ...810..135O} and which we endorse as the preferred approach to analyzing these data), for a given combination of model parameters such as $M$ and $R$ one determines the probability of the {\it data} given the {\it model} prediction using values of these parameters.  Hence, although the expected value of $\alpha$ given a set of parameter values can never exceed 1/8, the existence of measurement errors means that the value of $\alpha$ derived from observational data {\it can} exceed 1/8.  Even if the derived value of $\alpha$ is highly improbable given the model expectations, the probability will not be zero and thus the analysis will yield a credible region for $M$ and $R$, after marginalization over nuisance parameters.

As noted above, the frequentist method can sometimes require exclusion of measured data from the analysis.  The extent to which measured data are excluded provides a check of whether the model is consistent with the data.  In contrast, Bayesian parameter estimation uses all the data but assumes the model is correct and therefore cannot speak to whether the model is a good description of the data.  In practice, analyses of burst fluxes and spectra have often mixed elements of both approaches.  Using a frequentist analysis, the constraint that $\alpha$ be $\leq 1/8$ means that the distribution of solutions {\it cannot} include the $\beta=1/4$ line. On the other hand, when a Bayesian analysis is performed, the $\beta=1/4$ line {\it can} be a solution, and indeed the ghost solution tends to push the location of the peak in the posterior probability distribution in the $M$--$R$ plane toward the $\beta=1/4$ line. To see this, consider the following examples.

Suppose that perfect (i.e., accurate, zero-uncertainty) measurements of $F_{TD,{\rm obs}}$ and $A$ are available, that the true (but unknown) mass and radius of the star are $M_0$ and $R_0$, that the resulting true value of $\alpha$ is $\alpha_0$, that the  star is at a distance $D_0$, and that the true opacity is $\kappa_0$. Suppose also that we have perfect knowledge of the color factor $f_c$, but only approximate knowledge of the distance and the opacity. For any values of the distance $D$ and opacity $\kappa$, the mass and radius values given by
\begin{equation}
M=M_0\left(D\over{D_0}\right)^3{\kappa\over{\kappa_0}}{R_0\over R}\\
\label{eq:Mexact}
\end{equation}
and
\begin{equation}
R^2={D^2\over{2D_0^2}}{R_0^2\over{1-2\beta_0}}\left[1\pm\left(1-8{D\over{D_0}}{\kappa\over{\kappa_0}}\alpha_0\right)^{1/2}\right]\;,\\
\label{eq:Rexact}
\end{equation}
when inserted in the model, yield {\it exactly} the observed values of $A$ and $F_{TD,{\rm obs}}$. This means that if there is any uncertainty in $D$ or $\kappa$, the model yields exact solutions (i.e., solutions that agree exactly with the observed values of of $F_{TD,{\rm obs}}$ and $A$) for all the $M$--$R$ pairs on a curved line segment in the $M$--$R$ plane whose extent is limited by the uncertainties in $D$ and $\kappa$, and by the requirement that $8(D/D_0)(\kappa/\kappa_0)\alpha_0$ be $\leq 1$. 

\begin{figure}
\resizebox{0.5\textwidth}{!}{
  \includegraphics{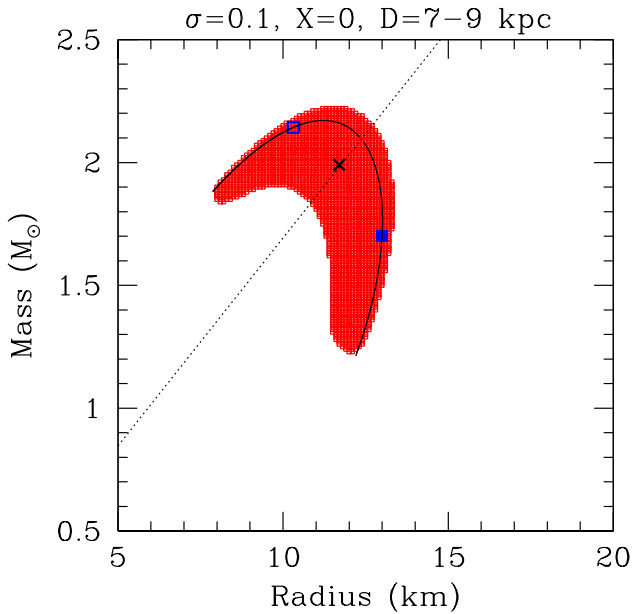}
}
\caption{Result of a joint analysis of five synthetic observed data sets, each of which was generated assuming $M=1.7~M_\odot$, $R=13$~km, $X=0$, and $D=8$~kpc. The $1\sigma$ statistical uncertainties in $F_{TD,{\rm obs}}$ and $A$ assumed when generating and analyzing the synthetic observational data were $0.1F_{TD,{\rm obs}}$ and $0.1A$. In this analysis, the prior for the distance was flat between $D=7$~kpc and $D=9$~kpc, i.e., the distance was assumed to be known fairly well. The solid blue square indicates the mass and radius assumed in generating the synthetic observed data, the open blue square indicates the ghost solution for the mass and radius, and the thick black cross indicates the peak of the posterior probability density over the $M$--$R$ plane. The red, horseshoe-shaped region is the most compact region that contains 68\% of the posterior probability. The black curve shows the loci of the exact solutions to Equation~(\ref{eq:Rexact}) for the distance interval used as the prior in the Bayesian analysis. Finally, the diagonal dotted line shows the relation $R=4GM/c^2$ ($\beta=1/4$). This figure illustrates two important issues encountered when using this model to analyze burst fluxes and spectra to determine the stellar mass and radius: even a small fractional uncertainty in the distance or the opacity (1)~produces a large region of nearly exact solutions and (2)~causes the posterior probability density to peak very close to the $R=4GM/c^2$ line. Note that the true values of $M$ and $R$ are within the 68\% credible region.}
\label{fig:MRsynth}
\end{figure}

This is illustrated in Figure~\ref{fig:MRsynth}, which shows the curve of exact solutions for a star with $M=1.7\,M_\odot$ and $R=13$~km. For this mass and radius, $R \approx 5.2GM/c^2$, $\beta=0.1931$, and $\alpha = 0.1185$. This figure shows that even if there is no uncertainty in the opacity and only a small fractional uncertainty in the distance, the curve of exact solutions spans a substantial fraction of the $M$--$R$ plane---the 12\% uncertainty in $D$ produces exact solutions that differ from the true solution by up to $\sim\,$30\% in $M$ and up to $\sim\,$50\% in $R$. Further analysis shows that if the uncertainties in $D$ and/or $\kappa$ are small enough, the curve of exact solutions develops a gap near $GM/Rc^2=4$. All the points on this curve are exactly consistent with the perfectly measured values of $A$ and $F_{TD,{\rm obs}}$, but only one point corresponds to the correct $M$--$R$ pair. The horseshoe shape of the curve is determined by the model, not the data.

Suppose now that the $F_{TD,{\rm obs}}$ and $A$ measurements have small fractional uncertainties. Figure~\ref{fig:MRsynth} shows the results of a Bayesian analysis of five synthetic observed data sets in which the statistical uncertainties in $F_{TD,{\rm obs}}$ and $A$ are 10\%. The resulting 68\% credible region has a ``broad horseshoe'' shape that tracks the curve of exact solutions that would be obtained if there were no uncertainties in the measured values of $F_{TD,{\rm obs}}$ and $A$. The maximum of the posterior probability density is on the $R=4GM/c^2$ ($\beta = 1/4$) line, which is where $\alpha = 1/8$, even though the relationship between the true radius and the true mass is $R\approx 5.2GM/c^2$.

As always in Bayesian analyses, prior probability distributions must be assigned to the model parameters.  In the analysis that produced the results shown in Figure~\ref{fig:MRsynth}, we assigned a flat prior in $M$ from 0.5-2.5~$M_\odot$, a flat prior in $R$ from 5-20~km, and a flat prior in $D$ from 7-9~kpc, and assumed that we had exact knowledge of $X$ (and thus of $\kappa$) and of $f_c$.  As noted in \cite{2015arXiv151007515S}, a flat prior in the stellar structure parameters $M$ and $R$ does {\it not} correspond to a flat prior in the equation of state parameters, such as $P$ and $\epsilon$.  If there is no other information available and the quantities of interest are $M$ and $R$, then a flat prior in those variables may be appropriate, whereas if the focus is on the equation of state, then flat priors in the equation of state parameters may be more appropriate.

Further analyses show that if the model is correct (i.e., the real system does not differ systematically from the model) and $F_{TD,{\rm obs}}$ and $A$ are measured perfectly but there are uncertainties in the distance or the composition, then there is always a local maximum in the posterior density at $\beta = 1/4$.

When $8\,(D/D_0)(\kappa/\kappa_0)\alpha_0 > 1$ for significant fractions of the $D$ and $\kappa$ priors, as will be the case if $\alpha_0$ is close to 1/8, the local maximum in the posterior probability density is likely to be close to the $\beta = 1/4$ line, even if the true $M$--$R$ pair is far from this line. More generally, if the analysis includes a spread of priors in $D$, $\kappa$, or (probably) $f_c$, the peak of the posterior probability density is likely to be close to the $\beta = 1/4$ line.

These results likely explain the otherwise puzzling fact that in the analyses reported in \cite{2015arXiv150505155O}, the location of the peak for all five bursting stars that have a single peak in their posterior probability density occurs at a point in the $M$--$R$ plane that is within 0.7\% of the $\beta = 1/4$ line.

It is important to note that the tendency of the probability density to peak close to the $\beta = 1/4$ line when this model is used does not necessarily mean that the resulting $M$ and $R$ estimates are significantly biased.  For instance, the true $(M,R)$ location can still be in the 68\% credible region, as it is in the example shown in Figure~\ref{fig:MRsynth}.  Our results show that $M$--$R$ estimates {\it can} be significantly biased if the true hydrogen mass fraction $X$ is at one end of the range considered in analysis (e.g., if the hydrogen mass fraction $X$ is zero, but $X = 0$ to $X = 0.7$ is considered in the analysis). The estimates can also be significantly biased if there is a systematic error in the model (e.g., if the entire stellar surface is not emitting uniformly). In these cases, the posterior probability density typically peaks at a point in the $M$--$R$ plane on the $\beta = 1/4$ line that is significantly different from the true mass and radius.

In conclusion, uncertainties (even moderate uncertainties) in distance, composition, or color factor, or systematic errors in the model, are the likely reasons why the peaks of the posterior probability densities found in~\cite{2015arXiv150505155O} are very close to the $\beta=1/4$ ($R=4GM/c^2$) line.

These peculiarities of the simple cooling model should be kept in mind when it is used to analyze burst data.

\subsection{Results obtained using the simple burst cooling model}

Recent representative radius estimates obtained by the three groups that have analyzed X-ray flux and color temperature data using the simple burst cooling model are $R=10.1$--11.1~km ($\,$$1\sigma$, i.e., likelihoods within a factor $e^{-1/2}$ of the peak likelihood)  for $M=1.5~M_\odot$ \cite{2015arXiv150505155O}; $R=10.4$--12.9~km (with 95\% confidence) for $M=1.4~M_\odot$ \cite{2013ApJ...765L...5S}; and $R=10.5$--12.8~km (with 95\% confidence) for $M=1.4~M_\odot$\cite{2015arXiv150906561N}.

The strong overlap of these radius ranges is remarkable and encouraging, especially because these analyses differ in their selection of which bursts to analyze (see the discussion in the following sections) and the strength of the nuclear physics priors used, including how causality is included as a constraint (see~\cite{2015arXiv151007515S} for a recent discussion). However, as we explained earlier, all of these methods share important simplifying assumptions. Most importantly, all assume that the entire stellar surface emits uniformly throughout the burst. One or more of these assumptions may have biased these estimates, in which case the true radius ranges would be shifted with respect to these ranges. At present, it appears plausible that all these radius estimates are somewhat smaller than the true radius, due to non-uniform emission from the stellar surface.

\subsection{Incompatibilities of the simple burst cooling model with burst data}

The compatibility of the simple burst cooling model with the observational data for a burst can be checked by comparing the value of $A$ derived from the data at touchdown (which we denote $A_{\rm TD}$) with the value of $A$ derived from the data during that part of the cooling tail when the flux is $\sim\,$70--10\% of the flux at touchdown (which we denote $A_{\rm tail}$)~\cite{2014MNRAS.445.4218K}. Spectra computed using model atmospheres consistently yield $f_c \approx 1.8$--1.9 at touchdown and $\approx\,$1.4--1.5 when the flux has declined to $\sim\,$70--10\% of the Eddington flux (see, e.g., \cite{2012A&A...545A.120S}). The simple cooling model assumes that the entire stellar surface emits uniformly at both times, and therefore predicts that $A$ is $\propto f_c^{-4}$. Hence, for the model to be compatible with the data, the ratio $A_{\rm tail}/A_{\rm TD}$ must lie between $\sim (1.8/1.5)^4\approx 2$ and $(1.9/1.4)^4\approx 3.4$.  

Figure~4 of \cite{2014MNRAS.445.4218K} shows that the area ratio $A_{\rm tail}/A_{\rm TD}$ for most bursts from most bursters is not compatible with the simple cooling model. In particular, bursts that occur in the soft spectral state and when the persistent flux exceeds $\sim\,$3\% of the Eddington flux far from the star generally have area ratios that are incompatible with this model (for an example of a well-observed burst in the soft spectral state that does not track the theoretical $f_c$ vs.\ flux curve, see Figure~1(d) of~\cite{2014MNRAS.445.4218K}). Appendix~A of~\cite{2014MNRAS.445.4218K} lists the area ratios (and their $\pm 1\sigma$ uncertainties) for each of the 240 bursts that were analyzed; taking the $\pm 1\sigma$ uncertainties in the area ratios into account, only 65 are compatible with the simple burst cooling model.

The results reported in~\cite{2015arXiv150505155O} for the bursts from five of the six stars that were analyzed are not compatible with the area ratio condition, taking the uncertainties of the area measurements into account (see Table~\ref{tab:normratio}). (In~\cite{2015arXiv150505155O}, the cooling tail is defined as the portion of the burst after the peak where the flux is 70--10\% of the touchdown flux.) The second column of Table~\ref{tab:normratio} shows the {\it upper} limits on $A_{\rm tail}$ obtained by increasing the value of $A_{\rm tail}$ listed in column~2 of Table~1 of \cite{2015arXiv150505155O} by one standard deviation (\cite{2015arXiv150505155O} does not give $A_{\rm tail}$ for the individual bursts from a given star, and $A_{\rm tail}$ is not available for the superburst from 4U~1820$-$30). The third column of Table~\ref{tab:normratio} shows the {\it lower} bounds on $A_{\rm TD}$ obtained by using the results shown in the upper right hand panels of Figures~3--8 of~\cite{2015arXiv150505155O} (these figures show the $1\sigma$ and $2\sigma$ contours for $A_{\rm TD}$ versus the color temperature for each PRE burst from a given star). The lower bound on $A_{\rm TD}$ listed for each star is the $2\sigma$ lower limit on $A_{\rm TD}$ for the burst from that star that has the smallest lower limit. Even when the values of $A_{\rm tail}$ and $A_{\rm TD}$ are pushed in the direction of consistency with the area ratio condition in this way, only the area ratio for EXO~1745$-$248 satisfies the condition. (The simple burst cooling model is incompatible with the {\it centroid} values of $A_{\rm tail}$ and $A_{\rm TD}$ obtained for EXO~1745$-$248.)

\begin{table}
\caption{Maximum area ratios from the data in \cite{2015arXiv150505155O}.  Normalizations are in units of (km/10~kpc)$^2$; a ratio between 2 and 3.4 is required for compatibility with the simple cooling model.}
\label{tab:normratio}       
\begin{tabular}{rrrr}
\hline\noalign{\smallskip}
Star & $A_{\rm tail,max}$ & $A_{\rm TD,min}$ & Max Ratio  \\
\noalign{\smallskip}\hline\noalign{\smallskip}
4U~1820$-$30 & 105.8 & 60 & 1.75 \\
SAX~J1748.9$-$2021 & 99.3 & 65 & 1.5 \\
EXO~1745$-$248 & 137.7 & 58 & 2.4 \\
KS~1731$-$260 & 103.9 & 90 & 1.15 \\
4U~1724$-$207 & 129.2 & 73 & 1.77 \\
4U~1608$-$52 & 358.3 & 230 & 1.56\\
\noalign{\smallskip}\hline
\end{tabular}
\end{table}

Another potentially troubling incompatibility between the simple burst cooling model and observational data was noted in~\cite{2010ApJ...722...33S}. As explained above, the observed data must give $\alpha \le 1/8$, otherwise the model yields a complex number for the stellar compactness. However, inserting the observed values of $F_{TD,{\rm obs}}$, $A$, and $D$ and the theoretically expected values of $\kappa$ and $f_c$ in Equation~(\ref{eq:def-alpha}) does not necessarily yield $\alpha \le 1/8$. Stated differently, the observed values of these quantities could produce $\alpha>1/8$, in which case the simple burst cooling model has no solution that is consistent with the data. Indeed, whether $\alpha$ is $\le 1/8$ could be considered a test of the validity of the model for the data being analyzed.  

The analysis in~\cite{2010ApJ...722...33S} shows that much of the burst data reported in~\cite{2010ApJ...719.1807G} does not satisfy the condition $\alpha \le 1/8$. The most striking example is the data on 4U~1820$-$30. The best estimates of $F_{TD,{\rm obs}}$, $A$, $D$, $\kappa$, and $f_c$ for this star that are reported in~\cite{2010ApJ...719.1807G} yield $\alpha=0.179$, which is substantially greater than 1/8. If one forms a probability space by combining the observed values of $A$, $F_{TD,{\rm obs}}$, and $D$ and their uncertainties with the computed theoretical range for $f_c$, only $\sim\,$$10^{-8}$ of the space is consistent with $\alpha\leq1/8$~\cite{2010ApJ...722...33S}. This indicates that the simple burst cooling model is strongly incompatible with these observations, taking the uncertainties in the observations into account.

Still another issue for the simple burst cooling model is that the peak of the bolometric flux during bursts from EXO~1745$-$248 (see Figure~3 of~\cite{2009ApJ...693.1775O}) and 4U~1724-307 (see Figure~4 of~\cite{2011ApJ...742..122S}) occurs at, rather than before, touchdown, contrary to what is expected in the model. In addition, a detailed study of all 16 normal thermonuclear X-ray bursts from 4U~1820$-$30 shows that the inferred emitting area in the tail is not constant~\cite{2013MNRAS.429.3266G}, contrary to what is assumed in the simple cooling model.

Thus, for many bursts the simple cooling model appears to be incompatible with the data. We now discuss efforts to address these incompatibilities.

\subsection{Efforts to address the incompatibilities of the simple burst cooling model with burst data}

The apparent incompatibilities of the simple burst cooling model with burst data have been addressed in three different ways: (1)~by analyzing only bursts that satisfy the area ratio test; (2)~by modifying the simple cooling model so that it is more consistent with the data, at the expense of introducing additional parameters; and (3)~by arguing that the limitations of the data preclude a meaningful test of the model, at least in those cases where the model appears incompatible with the data.

Advocates of the cooling tail method use the first approach, i.e., they analyze only bursts that satisfy the area ratio condition~\cite{2011A&A...527A.139S,2011ApJ...742..122S,2012A&A...545A.120S,2013IAUS..291..145P,2014MNRAS.442.3777P,2015arXiv150906561N,2015arXiv151006962S}. Although requiring a burst to meet this condition is an important step, it does not guarantee that the burst is accurately described by the simple cooling model. If in addition the observed path of the burst tail in the flux--color temperature plane follows the path predicted by the simple cooling model, this provides additional confidence in the use of the model. This information is available for some bursts. An example is the observed cooling track of the PRE burst from 4U~1608$-$52 plotted in panel (b) of Figure~1 in~\cite{2014MNRAS.445.4218K}. The $f_c$ vs.\ flux curve predicted by the simple cooling model follows the data closely, although it apparently does not give a formally good fit. A more comprehensive approach would be to do a sequence of fits of the spectra predicted by detailed model atmosphere calculations (see, e.g.,~\cite{2012A&A...545A.120S}) to the observed spectra of cooling bursts, and select for use in estimating $M$ and $R$ only bursts whose evolution is well fit, in the statistical sense, by these models (J.~N{\"a}ttil{\"a} et al., in preparation).

An example of the second approach, i.e., modification of the simple cooling model, is one of the approaches used in~\cite{2010ApJ...722...33S} and~\cite{2014EPJA...50...40L}. In some of the analyses reported in these papers, the radius of the photosphere at touchdown (i.e., the moment when the color temperature peaks) is allowed to be larger than the quiescent stellar radius (in the analysis in~\cite{2014EPJA...50...40L}, the photospheric radius at ``touchdown'' is formally allowed to be infinitely large). Although~\cite{2010ApJ...722...33S} and subsequent works in this vein find that some of the tension between the simple cooling model and the data can be eased by relaxing the requirement that the radius of the photosphere at touchdown be the radius of the quiescent stellar surface, a photospheric radius that remains significantly larger than the quiescent stellar radius is not physically plausible (the flux from the stellar surface would have to remain within $\sim\,$$10^{-4}$ of the Eddington flux during the measurements, in order for the atmosphere to remain expanded by $\sim\,$1~km). Analysis of high-precision measurements of the spectra of the 4U~1820-30 superburst, which is the only data set that can be analyzed with the required precision, rules out an atmosphere with a persistently large ($\gtorder\,$1~km) scale height with very high statistical confidence~\cite{2013IAUS..290..101M}.

Other possible modifications of the simple burst cooling model could include allowing the burst emission to come from only a fraction of the full stellar surface (see~\cite{2013IAUS..290..101M}), adding emission produced by accretion that continues during the burst, e.g., from a spreading accretion layer, or including reflection of the burst emission by the accretion disk (see \cite{2014MNRAS.445.4218K}). Burst radiation is expected to significantly affect the accretion flow, increasing or decreasing the accretion rate compared with the pre-burst rate and changing the spectrum of the accretion-powered emission~\cite{1992ApJ...385..642W,1996ApJ...470.1033M,1999ApJ...512L..35Y,2003A&A...399..663K,2004ApJ...602L.105B,2005ApJ...626..364B,2011A&A...534A.101C,2011A&A...525A.111I,2012PASJ...64...91S,2013ApJ...767L..37D,2013ApJ...772...94W,2014A&A...567A..80P}.

Assessing the third approach, i.e., arguments that the limitations of the data preclude a meaningful test of the model, which was taken in, e.g., \cite{2015arXiv150902924O}, requires a more detailed discussion.

It has been suggested~\cite{2015arXiv150902924O} that the apparent inconsistencies between the measured values of $A_{\rm tail}/A_{\rm TD}$ and the values predicted by the simple cooling model are caused by the rapid evolution of the fluxes and spectra of bursts near touchdown. In particular, in~\cite{2015arXiv150902924O} it was argued that $A$ is larger than $A_{\rm TD}$ before touchdown, because the photosphere is then larger than the quiescent stellar radius, and that even shortly after touchdown the measured value of $A$ is larger than $A_{\rm TD}$ because $A \propto f_c^{-4}$ and $f_c$ decreases rapidly after touchdown, as the flux decreases. The data near touchdown necessarily includes data from before and after the exact moment of touchdown. Therefore, it was argued, measurements of $A_{\rm TD}$ will be biased toward values larger than the true value, making the area ratio appear incompatible with the simple cooling model.

Using 0.25-second time bins, the fractional change in the flux between the touchdown bin and the adjacent bins was usually found to be 5--20\%, while the fractional change in $A$ between the touchdown bin and the adjacent bins was found to be 20--80\%~\cite{2015arXiv150902924O}. Thus, it is more difficult to accurately measure $A_{\rm TD}$ than $F_{\rm TD}$. It is nonetheless important to measure $A_{\rm TD}$, because the area ratio test is one of the few ways that the simple burst cooling model can be checked. The concerns raised in~\cite{2015arXiv150902924O} can be investigated in two ways: (1)~by evaluating how large an error is usually made in estimating $A_{\rm TD}$, and (2)~by considering other effects that would be predicted by this explanation for the apparent discrepancy between the measured area ratio and the ratio predicted by the simple cooling model.

We begin by evaluating the likely size of the error made in estimating $A_{\rm TD}$. It is important to note that comparing the values of $F_{\rm TD}$ and $A_{\rm TD}$ measured at touchdown with the values measured in the two adjacent time bins exaggerates the biases in the values of $F_{\rm TD}$ and $A_{\rm TD}$ measured at touchdown. The reason is that $A$ is expected to have a minimum at touchdown, which means that the rate of change of $A$ at touchdown is small. Assuming that $A$ varies quadratically near touchdown, the expected bias in the value of $A$ measured in the touchdown bin is several times less than the difference between the value of $A$ measured in the touchdown bin and the values of $A$ measured in the adjacent time bins. Even if the variation of $A$ near touchdown is V-shaped, the bias in the value measured at touchdown is 2--3 times smaller than the difference between the value in the touchdown bin and the adjacent bins, depending on the location of the actual moment of touchdown within the touchdown bin.

The suggestion made in~\cite{2015arXiv150902924O} can also be checked by measuring $A$ using a time resolution finer than 0.25~seconds. If the suggestion is correct, the value of $A_{\rm TD}$ measured using smaller time bins should be much less than the value measured using 0.25-second bins. However, work in progress (F.~Garcia, private communication) suggests that this does not explain the incompatibility: when the five bursts from 4U~1820$-$30 that were previously analyzed in this context are re-analyzed using data with time resolutions of 0.125~s and 0.0625~s, the resulting area ratios are still strongly inconsistent with the simple burst cooling model. Additional high-time-resolution analyses of more bursts near touchdown would help in evaluating this hypothesis.

Next, we look for other effects that would be expected if the evolution of the burst flux and spectrum near touchdown is biasing the estimate of $A_{\rm TD}$ to values large than the true value. If this is occurring, then if results for different bursts from the same star are compared, the estimate of $A_{\rm TD}$ should be smaller when the flux supposedly measured at touchdown is larger, because the relevant color correction factor should be larger.  Because the inferred value of $A$ scales as $f_c^{-4}$, this should be a large effect, and the anticorrelation should therefore be strong. If, for example, the color factor is 1.8 at the supposed moment of touchdown of one burst (when, say, the flux from the entire stellar surface is the Eddington flux) but 1.6 at the supposed moment of touchdown of another burst (perhaps because the flux at the supposed moment of touchdown for that burst is actually 10\% less than the Eddington flux), then the estimate of $A_{\rm TD}$ for the former burst should be 1.6 times smaller than the estimate for the latter burst.

However, at least for the bursts analyzed in~\cite{2015arXiv150505155O}, the inferred values of $A_{\rm TD}$ do not show the strong anticorrelation with the measured touchdown flux predicted by this hypothesis. Using the results presented in panel~(b) of each of Figures 3--8:

\begin{enumerate}

\item 4U~1820-30 has a fairly tight collection of fluxes, so these measurements do not provide much information about the relationship of the measured touchdown flux and the area normalization at touchdown in this star.

\item SAX~J1748.9$-$2021 has two bursts that were included in this analysis. One clearly has a higher touchdown flux than the other (by about 20\%).  The burst with the higher touchdown flux has, if anything, a {\it larger} area normalization, although the two areas are consistent with one another within the statistical uncertainties.

\item EXO~1745$-$248 has two bursts with contours that almost coincide, so these measurements do not provide much information about the relationship of the measured touchdown flux and the area normalization at touchdown in this star.

\item KS~1731$-$260 has two bursts with nearly identical touchdown fluxes but substantially different area normalizations. If anything, the burst with the higher touchdown flux has a slightly larger area normalization.

\item 4U~1724$-$207 has two bursts. The touchdown flux and the area normalization at touchdown are anticorrelated, but the strength of the anticorrelation is much smaller than what would be expected. The lower touchdown flux is about 75\% of the higher touchdown flux, so the area normalization of the burst with the lower flux should be about 300\% of the area normalization of the burst with the higher flux. Instead, it is only about 140\% of the area normalization of the burst with the higher flux.

\item 4U~1608$-$52 has three bursts. The bursts with higher touchdown fluxes tend to have larger area normalizations, rather than the reverse.

\end{enumerate}

\noindent  Thus, of the four stars that can be used to evaluate the hypothesis that the evolution of the burst flux and spectrum near touchdown is biasing the estimate of $A_{\rm TD}$, none quantitatively match the behavior expected given this hypothesis, and at least two do not qualitatively match the expected behavior.

What about the large number of bursts that give estimates of $\alpha$ that exceed 1/8 and hence are inconsistent with the simple cooling model? It has been suggested that this apparent inconsistency results from using inappropriately small measurement uncertainties when analyzing the burst data~\cite{2015ApJ...810..135O}. Specifically, it was argued that rather than using the statistical uncertainties in the individual measurements of $A$ and $F_{TD,{\rm obs}}$, which are typically 3--5\% (see, e.g., Table~2 of~\cite{2010ApJ...719.1807G}), when estimating $M$ and $R$ the uncertainties in $A$ \cite{2012ApJ...747...76G} and $F_{TD,{\rm obs}}$ \cite{2012ApJ...747...77G} should instead be set to $\sim 10$\%, because independent measurements of these quantities using different bursts from the same star typically vary by $\sim 10$\%. This variation is not due to  uncertainties in the detector calibration: the Proportional Counter Array (PCA) onboard {\it RXTE}, which is the detector used to collect most of the burst data that have been analyzed, is calibrated to better than 1\% \cite{2006ApJS..163..401J}.

If the measured values of parameters that are fixed in the model being evaluated are found to vary by amounts greater than the statistical uncertainties of the measurements, this indicates that the model in question is incorrect or incomplete. Such variations should induce caution, because using an incorrect model can bias the results inferred using the model. For example, if the temperature of the burst emission varies over the stellar surface (as was plausibly suggested in~\cite{2015arXiv150902924O}), fitting a uniform-temperature model to burst data will underestimate the emitting area. This will happen systematically, so even if the temperature variation changes from burst to burst, the radius estimate made using the model will be biased toward a value smaller than the true value. It is possible that there could be similar effects on the touchdown flux. Even if the measured fluxes or area normalizations have Gaussian distributions with outliers that are removed, the {\it spread} in the measured values cannot in general be treated as a {\it statistical} uncertainty if it exceeds the measurement errors. In particular, because systematic errors do not average out in the same way that statistical errors do, analyzing a large number of bursts from the same star cannot reduce such errors. Systematic errors cannot be treated in the same way as statistical uncertainties.

In~\cite{2015ApJ...810..135O}, it was suggested that the situation noted in~\cite{2010ApJ...722...33S}, namely, that only a small fraction (in the case of 4U~1820$-$30, only $\sim\,$$10^{-8}$) of the probability space used in estimating $M$ and $R$ with the simple cooling model is compatible with the model, is caused by the scatter in the measured values of $A$ and $F_{TD,{\rm obs}}$. In~\cite{2015ApJ...810..135O}, this idea was tested by generating synthetic data for a $1.7~M_\odot$ star with a radius of 10~km and a zero hydrogen mass fraction ($X=0$) at a distance $D$ of 5~kpc. The color factor $f_c$ was fixed at 1.35. The data were then analyzed using a flat prior on $X$ from $X=0$ to $X=0.7$ and a flat prior on $D$ from $D_{\rm min} = 4.9$~kpc to an upper limit $D_{\rm max}$ that ranged from 5.1~kpc to 15~kpc, depending on the case considered. For this particular example, the fraction of the trials that are compatible with the model is usually much greater than $10^{-8}$, but does have a small probability of being as small as the $\sim\,$$10^{-8}$ fraction found in~\cite{2010ApJ...722...33S} for 4U~1820$-$30, if the uncertainty in $A$ and $F_{TD,{\rm obs}}$ assumed when analyzing the synthetic data is five times larger than the uncertainty used when generating the synthetic data, especially if $D_{\rm max}$ is substantially larger than the actual value of $D$. This does not mean, however, that such a small fraction is what would typically be found when analyzing  burst data, because the parameter values chosen for this example {\it guarantee} that the fraction of cases that will be consistent with the model will be small.

Choosing $M=1.7~M_\odot$ and $R=10$~km means that the true value of $\beta$ is 0.25105, and hence that the true value of $\alpha=\beta(1-2\beta)$ is 0.1249978. This value of $\alpha$ is so close to the $1/8=0.125$ maximum value of $\alpha$ that is compatible with the simple cooling model (differing from it by less than 0.002\%) that measured values of $\alpha$ that are even slightly larger than the true value will be incompatible with the model. As Equation~(\ref{eq:def-alpha}) shows, if the value of $\kappa$ that is chosen for a trial is even slightly larger than the true value, or if the value of $D$ that is chosen for a trial is even slightly larger than the true value, the resulting value of $\alpha$ will be larger than 0.125, and hence incompatible with the model. Given that the value of $X$ assumed in generating the synthetic data for the example was 0 but the search interval was chosen to range from $X=0$ to $X=0.7$, almost the entire range of hydrogen mass fractions that were considered were guaranteed to produce unphysical solutions. The star 4U~1820$-$30, which has the lowest consistency probability of all currently analyzed systems and thus poses the strongest challenge to the simple cooling model, is in an ultracompact binary system and is therefore known to have $X=0$, so for this system there is no reason to search over $X$. Similarly, because the value of the distance $D$ assumed in generating the synthetic data was 5~kpc but the search interval for $D$ used in analyzing the data ranged from 4.9~kpc up to $D_{\rm max}$, with $D_{\rm max}$ as large as 15~kpc, almost the entire range of $D$ that was considered is guaranteed to produce a value of $\alpha$ larger than the maximum value that is compatible with the model. Thus, the choices made in~\cite{2015ApJ...810..135O} for $M$ and $R$, coupled with the choices made for the prior probability distributions in $X$ and $D$ compared to the values used in generating the synthetic data, produce a misleading impression. The concern raised in~\cite{2010ApJ...722...33S} about 4U~1820$-$30 and similar stars remains valid: a high fraction of the burst data that has been gathered is incompatible with the simple burst cooling model that is being used to estimate the stellar mass and radius. 

In summary, analysis of the evolution of the X-ray flux and spectrum during X-ray bursts is a promising method for estimating $M$ and $R$. Using this method, radii $\sim\,$11--13~km are commonly found for $M \approx 1.4$--$1.5~M_\odot$. However, the inconsistencies between the predictions made by the simple cooling model currently being used and the burst data, and the possibility that important aspects of the burst emission (such as non-uniform emission over the stellar surface) are currently not included in the model, mean that the $M$ and $R$ estimates made using this approach may be affected by significant systematic errors.

Very high precision future data could yield qualitatively new information that will make it possible to check the burst models and correct for these systematic errors.  For example, if only part of the surface is emitting actively, the rest of the surface will be cooler.  An instrument with good low-energy response might be able to detect that cool portion of the star and thus make it possible to estimate the fraction of the star that is hot.  If observer inclinations can be measured for these systems, it may be possible to assess the likelihood of disk obscuration or estimate the contribution due to reflection if the observations yield information about disk warps and aspect ratios.  In addition, if an instrument with excellent low-energy response can determine the luminosity when the atmosphere is highly expanded (radius $\gtorder 10\times$ the stellar radius) during a strong PRE burst, the redshift from the thermalization radius will be negligible and the mass of the star can therefore be measured directly.  This information could then be combined with the spectra and fluxes at touchdown to infer $R$.  Even if the thermalization radius is not so highly expanded, accurate measurements of the flux and spectrum could make it possible to determine $M/R$ (see the discussion of this possibility in \cite{1987MNRAS.226...39S}).  This information could then be combined with data from the tail to obtain $M$ and $R$ separately.  Future missions that could provide very high-resolution spectra with high time resolution could constrain or determine the extent of any absorption, reflection, or obscuration of the burst emission, or contamination of the flux by other emission from the system.

\section{Mass and Radius Measurements Using X-ray Oscillations Produced by Stellar Rotation}

If a neutron star has a hotter region (``hot spot") on its surface and gas in the hot spot rotates around the star, a distant observer will see an energy-dependent waveform whose properties are affected by special relativistic effects such as Doppler shifts and aberration, and general relativistic effects such as light deflection (for various levels of approximation for treating these effects, see \cite{1983ApJ...274..846P,1998ApJ...499L..37M,2002ApJ...566L..85B,2003MNRAS.343.1301P,2007ApJ...663.1244M,2014ApJ...792...87P}).  The magnitudes of these effects depend on the mass and radius of the star, and thus the mass and radius can be measured by carefully analyzing such waveforms.  

Hot spots are observed to rotate with frequencies that are the same as the rotational frequency of the underlying star or very close to it.  In practice, a rotational frequency of hundreds of Hz is required to measure $M$ and $R$ with uncertainties of a few percent.  Consequently, this method provides the strongest constraints when it is used to analyze X-ray observations of accretion-powered millisecond pulsars (\cite{2003MNRAS.343.1301P,2006MNRAS.373..836P,2008ApJ...672.1119L,2009ApJ...691.1235L,2011ApJ...742...17L,2011ApJ...726...56M}; see \cite{2012arXiv1206.2727P} for an observational description of these stars), rotation-powered millisecond pulsars \cite{2000ApJ...531..447B,2007ApJ...670..668B,2008ApJ...689..407B,2013ApJ...762...96B}, or the millisecond brightness oscillations observed during some thermonuclear X-ray bursts (\cite{1997ApJ...487L..77S,1998ApJ...499L..37M,2001ApJ...546.1098W,2013MNRAS.433L..64A,2013ApJ...776...19L,2014ApJ...787..136P,2015ApJ...808...31M,2015ApJ...811..144B}; see \cite{2012ARA&A..50..609W} for a discussion of the observations).  

Our description of this method draws on \cite{2013ApJ...776...19L} and \cite{2015ApJ...808...31M}. We focus here on analysis of the waveforms of X-ray burst oscillations and rotation-powered X-ray pulsars because the complications created by pulse profile variability (see, e.g., \cite{2008ApJ...675.1468H}) and Comptonization by the hot accreting gas may introduce significant systematic errors when analyzing the waveforms of accretion-powered millisecond pulsars.  On a more optimistic note, the pulse profile of SAX~J1808-3658 can be relatively stable \cite{2009MNRAS.400..492I,2009ApJ...706L.129P}, and its geometry is relatively well-known.  It is also possible that polarization missions can help obtain mass-radius constraints on these stars (J.~Poutanen, private communication).

The standard model for the waveform produced by a single spot on the surface of a rotating star has seven main parameters: (1)~the gravitational mass and (2)~circumferential radius of the star, (3)~the colatitude of the observer, (4)~the colatitude of the center of the hot spot, (5)~the angular radius and (6)~the surface comoving temperature of the hot spot (which is assumed to be a uniformly emitting circular area), and (7)~the distance to the star. This single-spot model appears adequate for analyzing the waveforms of X-ray burst oscillations, which is a key goal of the proposed {\it LOFT} \cite{2012ExA....34..415F} and {\it AXTAR} \cite{2011arXiv1109.1309R} missions. In contrast, the analysis of waveforms produced by thermal X-ray emission from the heated polar caps of rotation-powered pulsars, which will be the focus of the accepted {\it NICER} mission \cite{2012SPIE.8443E..13G}, may require inclusion of a second spot (see, e.g., \cite{2009ApJ...703.1557B,2013ApJ...762...96B}).

Auxiliary parameters must also be included in the analysis.  One such parameter is the time at which the emission peaks during the rotational period.  A parameter must also be included in the waveform model for each energy channel in the observing instrument, to represent the number of background counts in each channel.  This takes into account any backgrounds that do not vary in a way that is commensurate with the rotational frequency of the star.  Possible sources of background counts include non-spot emission from the stellar surface, other sources in the field of view, the instrumental background, and, in the case of X-ray burst sources, emission from the accretion disk that surrounds the star.  For burst oscillations, the background during a burst can differ significantly from the background before or after the burst \cite{1992ApJ...385..642W,1996ApJ...470.1033M,1999ApJ...512L..35Y,2003A&A...399..663K,2004ApJ...602L.105B,2005ApJ...626..364B,2011A&A...534A.101C,2011A&A...525A.111I,2012PASJ...64...91S,2013ApJ...767L..37D,2013ApJ...772...94W,2014A&A...567A..80P}, so it is essential to allow the background to have an arbitrary total count rate and energy spectrum.  Fortunately, it is straightforward to include such a background in a Bayesian analysis \cite{2013ApJ...776...19L,2015ApJ...808...31M}.  

Some of the parameters in the waveform model are partially degenerate with each other.  For example, in many circumstances changes in the observer's inclination to the stellar spin axis can have effects on the waveform that are similar to the effects produced by changes in the colatitude of the hot spot. However, with data of the quality that is anticipated from the {\it NICER} mission and that would be obtained from the {\it LOFT} or {\it AXTAR} missions, it is possible to break these degeneracies and determine $M$ and $R$ to within a few percent, the precision needed to provide useful guidance for developing models of the properties of cold neutron-rich matter at supranuclear densities. In particular (summarizing results from \cite{2013ApJ...776...19L} and \cite{2015ApJ...808...31M}):

\begin{enumerate}

\item The primary figure of merit for determining the precision that can be achieved in measuring $M$ and $R$ is ${\cal R} \equiv N_{\rm osc}/\sqrt{N_{\rm tot} }= f_{\rm rms}\sqrt{N_{\rm tot}}$.  Here $N_{\rm osc}$ is the number of X-ray counts in the  oscillating component of the waveform, $N_{\rm tot}$ is the total number of X-ray counts, and $f_{\rm rms}$ is the fractional root-mean-square amplitude of the oscillation.  For fixed values of the other parameters in the waveform model, the precisions of $M$ and $R$ estimates scale as ${\cal R}^{-1}$.

\item Significant harmonic structure (i.e., detectable overtones of the rotational frequency) is necessary to achieve precise measurements of $M$ and $R$.  This requires spot rotational frequencies $\nu_{\rm rot} \gtorder 300$~Hz and spot and observer inclinations within $\sim 30^\circ$ of the rotational equator.  Then ${\cal R} \gtorder 400$ is sufficient to achieve uncertainties in $M$ and $R$ that are $\ltorder 10$\%.  In a highly favorable case in which $\nu_{\rm rot}=600$~Hz and $\theta_{\rm obs}=\theta_{\rm spot}=90^\circ$, achieving ${\cal R}=400$ would make it possible to determine $M$ and $R$ with uncertainties of less than 3\% \cite{2015ApJ...808...31M}.

\item If the modulation fraction is \mbox{$\sim\,$10\%} \cite{2014ApJ...792....4C}, achieving ${\cal R}=400$ requires \mbox{$\sim\,$10$^7$} total counts.  For the thermal X-ray oscillations produced by rotation-powered pulsars, collecting \mbox{$\sim\,$10$^7$} total counts requires long observations (days to weeks, using {\it NICER}). For X-ray burst oscillations, this requires combining data from many bursts (25--50, using {\it LOFT}).  In the latter case, data from burst rises, peaks, and tails can be used: combining data from different bursts or from different parts of the same burst is straightforward using Bayesian joint analysis techniques (see Appendix~B of \cite{2013ApJ...776...19L}).

\item For some individual stars, it may be possible to constrain some of the parameters in the waveform model independently of the waveform-fitting process.  This would reduce the uncertainties in $M$ and $R$; \cite{2013ApJ...776...19L} find that independent knowledge of $\theta_{\rm obs}$ would reduce these uncertainties substantially, whereas independent knowledge of $D$ would reduce them only modestly.

\end{enumerate}

The currently available data do not provide large enough ${\cal R}$ values for any neutron stars that tight constraints can be placed on their masses or radii using waveform fitting. The main reason is that previous and current instruments do not provide enough counts. The best current analysis~\cite{2013ApJ...762...96B} concludes that the rotation-powered pulsar PSR~J0437-4715, which is not accreting and has a mass of $M=1.44\pm 0.07~M_\odot$~\cite{2016MNRAS.455.1751R}, has a radius $>10.7$~km at the $3\sigma$ confidence level, assuming its mass is $1.44\,M_\odot$ \cite{2013ApJ...762...96B}. This analysis also assumes that there are two spots and that the spectrum and beaming pattern of the radiation from the spots is correctly described by radiation from a hydrogen atmosphere in which the energy is being deposited at a very large optical depth. Finally, it assumes that both spots are identical, although not necessarily antipodal, and that the they have a hot core inside a cooler annulus.

As noted in previous sections, a critical question is whether systematic errors dominate the statistical errors in the estimates of neutron star masses and radii. A reason to be cautiously optimistic about the waveform fitting method is that none of the studies carried out to date using synthetic waveform data have identified a case in which (1)~the data are fit well in a statistical sense using the standard waveform model and (2)~$M$ and $R$ appear to be tightly constrained, but (3)~the mass and radius estimates derived by fitting the waveform to the data are significantly biased relative to the values used in generating the data.  That is, if a synthetic waveform is generated using assumptions different from the assumptions used in analyzing the waveform, the resulting mass and radius estimates are consistent with the values of the mass and radius used in generating the waveform if the fit is statistically good and the constraints are tight.  Deviations from the standard model that have been explored include differences in the energy spectrum, the beaming pattern from the stellar surface, the shape of the hot spot, and the temperature profile of the spot \cite{2013ApJ...776...19L,2015ApJ...808...31M}.  More work is needed, but at present the waveform-fitting method appears to suffer less from systematic errors than other methods that provide simultaneous estimates of the stellar mass and radius.  

Ultimately, it will be useful to measure the masses and radii of stars for which multiple methods can be used. For example, J.~Kajava (private communication) notes that Aql~X-1 produces PRE bursts, burst oscillations, and intermittent accretion-powered pulsations, and also has quiescent periods during which its cooling emission can be observed.

\section{Other Possibilities for Future Mass and Radius Measurements}

Additional constraints on dense matter via electromagnetic observations could come from radio observations using the Square Kilometer Array, such as measurements of pulsar precession in double neutron star systems that could help determine the moment of inertia to $\sim\,$10\% (see, e.g., \cite{2015aska.confE..43W}). If neutron stars are detected with spin frequencies considerably larger than the current maximum of 716~Hz~\cite{2006Sci...311.1901H}, this could set a useful lower limit on the average density of the star. If oscillation modes of neutron stars can be identified unambiguously (e.g., as quasi-periodic oscillations in the tails of magnetar superbursts~\cite{2005ApJ...628L..53I,2005ApJ...632L.111S,2006ApJ...653..593S,2006ApJ...637L.117W}), the detailed properties of these modes will provide information about the dense matter in the cores or crusts of the stars~\cite{2005ApJ...632L.111S,2006ApJ...653..593S,2007MNRAS.379L..63W,2007MNRAS.374..256S,2007Ap&SS.308..581S,2008MNRAS.385L...5S,2009PhRvL.103r1101S,2012MNRAS.421.2054G}.

In terms of future methods, most attention has been focused on the prospects that gravitational waves from coalescing compact binaries will yield mass and radius constraints that are entirely independent of the constraints derived from electromagnetic observations. The potential of this method has been demonstrated by the recent detection of a burst of gravitational radiation by {\it LIGO}~\cite{2016arXiv160203837T,2016arXiv160203846T} and analysis of the waveform, which agrees quantitatively with numerical computations of the waveform produced by the inspiral and merger of two massive black holes in a binary system~\cite{2016arXiv160203837T,2016arXiv160203846T}.

The waveforms from the inspiral and merger of two neutron stars or a neutron star and a black hole bear the imprint of the tidal interactions of the stars (e.g., \cite{2009PhRvD..79l4033R,2010PhRvD..81l3016H}).  Although many high-precision numerical simulations are still needed, early indications are that analytical models of the tidally-influenced waveform are sufficiently accurate to perform reliable parameter estimation \cite{2015PhRvL.114p1103B}.  The detection of several to tens of such events may allow discrimination between soft, medium, and hard equations of state, although it could be important to have good prior knowledge of the distribution of neutron star masses \cite{2015PhRvD..92b3012A}.

It may be possible to place useful constraints on the maximum mass of neutron stars by combining gravitational-wave and electromagnetic observations of short gamma-ray bursts.  These bursts are thought to be produced by the merger of two neutron stars or a neutron star and a black hole.  If two neutron stars merge, then \cite{2014ApJ...788L...8M} (see also \cite{2002MNRAS.336L...7R}) argue that the merged remnant must collapse within \mbox{$\ltorder\,$0.1~s}, otherwise the baryonic wind driven by neutrinos will delay and lengthen the bursts beyond the few tenths of a second duration that is observed.  This line of argument implies that a successful burst requires that the total baryonic mass of the two original neutron stars exceed the maximum that can be sustained by a uniformly rotating star \cite{2015ApJ...812...24F,2015ApJ...808..186L} (note that if one believes that the \mbox{$\sim\,$100~s} X-ray plateaus seen in some short bursts require that the rotating star be stable, the limit goes in the other direction \cite{2013PhRvD..88f7304F}).  \cite{2015ApJ...808..186L} find that if the masses of neutron stars that merge to produce short gamma-ray bursts are comparable to those of the neutron stars we see in our Galaxy, then the maximum mass of a nonrotating neutron star is $\sim\,$2.05--2.2~$M_\odot$, which is quite close to the $2.01~M_\odot$ maximum mass currently observed.  Obtaining more reliable constraints will require identification of individual short bursts with specific gravitational-wave events. Because the gamma rays from these bursts are tightly beamed and are therefore seen by only a small fraction of observers, such associations are likely to require signatures in other electromagnetic bands that can be seen over a much broader solid angle (e.g., optical to infrared ratios in kilonovae \cite{2014MNRAS.441.3444M,2015MNRAS.450.1777K} or possibly scattered X-rays \cite{2015ApJ...809L...8K}).

\vspace{20pt}

\noindent  We have benefitted from stimulating conversations with Zaven Arzoumanian, David Blaschke, Slavko Bogdanov, Deepto Chakrabarty, Federico Garc\'{\i}a, Sebastien Guillot, Tolga G\"uver, Craig Heinke, Jari Kajava, Jim Lattimer, Mariano M\'{e}ndez, Sharon Morsink, Joonas N\"attil\"a, Feryal \"Ozel, Juri Poutanen, Dimitrios Psaltis, Bob Rutledge, Andrew Steiner, Valery Suleimanov, Anna Watts, and Guobao Zhang.

\bibliography{epja}

\end{document}